\documentclass[journal=nalefd,manuscript=article]{achemso}
\usepackage{graphicx}
\usepackage{amsmath}
\usepackage{amssymb}
\usepackage{bm}

\author{Jos\'e L. Movilla} 
\affiliation{Dept. d'Educaci\'o i Did\`actiques Espec\'ifiques, Universitat Jaume I, 12080, Castell\'o, Spain}
\author{Josep Planelles}
\affiliation{Departament de Qu\'{\i}mica F\'{\i}sica i Anal\'{\i}tica,
Universitat Jaume I, E-12080, Castell\'o de la Plana, Spain}
\author{Juan I. Climente}
\affiliation{Departament de Qu\'{\i}mica F\'{\i}sica i Anal\'{\i}tica,
Universitat Jaume I, E-12080, Castell\'o de la Plana, Spain}
\email{climente@uji.es}

\date{\today}

\title{ Dielectric Confinement Enables Molecular Coupling in Stacked Colloidal Nanoplatelets}


\begin{document}

\begin{abstract}

We show theoretically that carriers confined in semiconductor colloidal nanoplatelets (NPLs) sense
the presence of neighbor, cofacially stacked NPLs in their energy spectrum. 
When approaching identical NPLs, the otherwise degenerate energy levels redshift and split, 
forming (for large stacks) minibands of several meV width.
Unlike in epitaxial structures, the molecular behavior does not result from quantum tunneling 
but from changes in the dielectric confinement.
The associated excitonic absorption spectrum shows a rich structure of bright and dark states, whose optical activity 
and multiplicity can be understood from reflection symmetry and Coulomb tunneling. 
We predict spectroscopic signatures which should confirm the formation of molecular states,
whose practical realization would pave the way to the development of nanocrystal chemistry based on NPLs.
\end{abstract}


\begin{tocentry}
\begin{center}
\includegraphics[width=8.25cm]{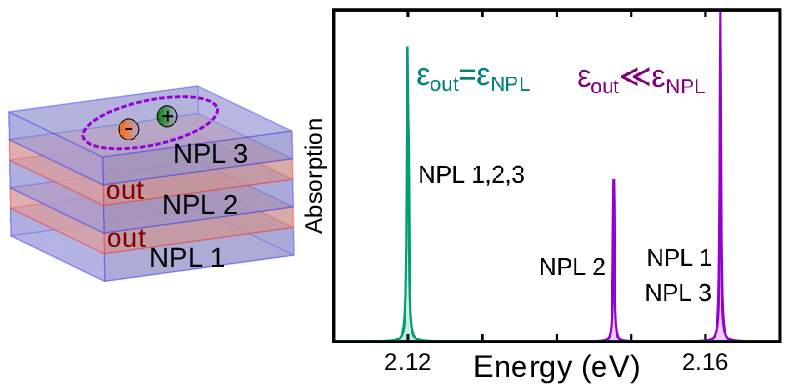} 
\end{center}
\end{tocentry}

\newpage

Semiconductor quantum dots synthesized via wet chemistry, also known as colloidal nanocrystals,
have reached a level of maturity that makes them suitable for existing and emerging optical and 
electronic technologies.\cite{KovalenkoACS}
A natural step for further progress is the controlled coupling of nanocrystals, by which the atomic-like 
energy levels of the dots interact with those of neighbors to form a molecular-like spectrum.
Current endeavor to develop this so-called ``nanocrystal chemistry'' largely relies on the design of coupled 
nanocrystal structures, from dimers\cite{TeitelboimACR,CuiXXX} to superlattices\cite{BolesCR,WhithamNM}, 
where quantum tunneling of charged carriers\cite{TadjinePCCP} (electrons, holes) or 
dipole-dipole interactions of neutral carriers\cite{CrookerPRL,RowlandNM,GuzelturkACS} (excitons) are exploited.

As a newcomer in the family of nanocrystals, colloidal NPLs have attracted much interest in the last years.\cite{LhullierACR,TyagiJPCL}
Their quasi-two-dimensional structure with precise (atomic monolayer) thickness control results in extremely bright and narrow fluorescent emission,
which makes them particularly promising for optoelectronic applications.\cite{KovalenkoACS,LhullierACR,TyagiJPCL,RowlandNM,GrimNN,ScottNN}
On the other hand, the potential of NPLs as building blocks for molecular superstructures with electronic coupling is still unclear.
Van der Waals attraction enables the self-assembly of cofacially stacked NPLs, where the inorganic semiconductor alternates 
with few-nm-thick layers of organic ligands.\cite{JanaLANG,LeeNL}
The stacked NPLs can be chosen to have identical thickness, thus constituting a nearly ideal homonuclear molecule,
with negligible size polydispersity.
However, the formation of electron or hole energy minibands driven by quantum tunneling 
--analogous to those in epitaxial quantum well superlattices\cite{QW_book}-- 
is inhibited because the ligands typically impose a high (2-4 eV) potential barrier.
The constitution of such energy minibands would be of great practical interest, 
as an additional degreee of freedom in the electronic structure design and to possibly combine 
the outstanding optical properties of individual NPLs with transversal transport properties 
of superlattices.

Recently, low temperature photoluminescence measurements of CdSe NPLs have revealed two well 
resolved emission peaks instead of one.\cite{TessierACS,AchtsteinPRL,YuACSami,ShornikovaNL,DirollNL} 
Diroll and co-workers showed that the relative intensity and energy splitting of the two peaks 
depend on the degree of stacking and inter-NPL distance, respectively.\cite{DirollNL}
 These observations suggest that inter-NPL excitations take place in stacked NPLs.
With quantum tunneling quenched by the organic spacer layer, the fundamental question arises of
what the origin of such interactions is, and how can they be controlled.

The goal of this work is to provide a theoretical framework to answer the above questions.
We calculate the electronic structure of electrons, holes and excitons in stacked NPLs. 
We shall see that molecular interactions are in fact enabled by the drastically different polarizability 
of NPLs and ligands, which make self-energy and Coulomb screening terms sensitive to the number, 
position and proximity of nearby NPLs. 
Beyond assessing on Ref.~\cite{DirollNL} observations, our results provide a general 
picture of how inter-NPL interactions operate in colloidal systems
and what experimental signatures should be sought in future experiments to confirm the formation of
molecular states driven by dielectric confinement.


Charged carriers are described with effective mass Hamiltonians,
\begin{equation}
H^j = \frac{\mathbf{p_\perp}^2}{2 m_\perp^j} + \frac{p_z^2}{2 m_z^j} + V_{bo}^j(\mathbf{r_j}) 
+ \Sigma(\mathbf{r_j}).
\label{eq:H1}
\end{equation}
\noindent where $j=e,h$ stands for electron and hole, respectively. 
$m_{z}^j$ ($m_\perp^j$) is the mass parallel (perpendicular) to the 
strong confinement direction of the NPL, $V_{bo}^j$ is the potential band-offset between the
NPL and the organic environment, and $E_\Sigma$ the self-energy potential arising from the 
dielectric mismatch between the two materials.\cite{RodinaJETP,Delerue_book,EvenPCCP,BenchamekhPRB}
\begin{figure}[h]
\includegraphics[width=8.5cm]{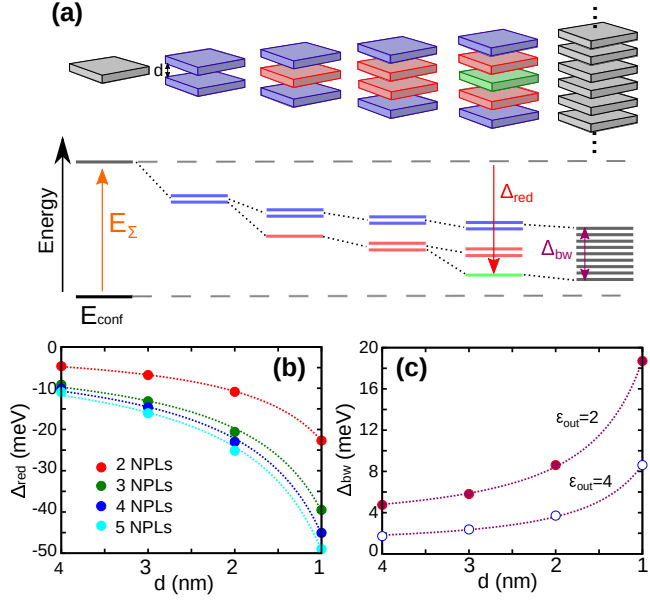}
\caption{
(a) Schematic energy spectrum of a single particle in stacks of NPLs. With increasing number of NPLs, $\Sigma$ gradually weakens, the energy levels redshift and degeneracy is lifted. 
Minibands are expected for periodic arrays of NPLs. 
(b) Ground state redshift for different number of NPLs,
as a function of the inter-NPL distance. Dots are calculated values, lines $1/d$ fits.
(c) Band width of a 5 NPL stack, for two outer dielectric constants. 
Dots are calculated values, lines $1/d^{3/2}$ fits.
}
\label{fig1}
\end{figure}

Since the ratio of NPL/ligand dielectric constants fulfills $\epsilon_{in}/\epsilon_{out} > 1$, 
a charge inside the NPL creates a surface polarization field of the same (opposite) 
sign on the inner (outer) side of the NPL.\cite{RodinaJETP,Delerue_book}
$\Sigma$ is then repulsive inside the NPL and attractive in between NPLs.
From this basic consideration, we can foresee the qualitative electronic structure of stacked NPLs,
which is shown in Figure \ref{fig1}(a).  
For a single NPL (left part of figure), assuming infinite potential barrier, 
the electron (or hole) ground state energy is given by $E_{conf} + E_\Sigma$. 
The former is the quantum confinement energy and the latter the correction coming from self-energy repulsion.
When two NPLs are considered, since tunneling is largely suppressed in colloidal systems, 
quantum confinement energy is again $E_{conf}$.
However, if the inter-NPL distance $d$ is short enough, a charge confined inside a NPL polarizes not only the host NPL but also
the neighbor one. The dielectric confinement is then weaker than for isolated NPLs, and $E_\Sigma$ decreases.
This gives rise to a redshift of the (doubly degenerate) ground state. 
For three NPLs, the same reasoning implies the redshift further increases. 
Moreover, in this case the redshift is more pronounced 
for the central platelet than for the terminal ones, because two nearby (first) neighbors weaken dielectric confinement
more efficiently than a first and a second neighbor. 
This asymmetry of dielectric environment splits the otherwise triply degenerate energy levels into a 
singlet (central NPL) and a doublet (two terminal NPLs). 
Similar trends hold for increasing number of NPLs, but the redshift saturates because the influence of distant NPLs 
vanishes. At the same time, the number of states keeps on increasing. 
As a result, the electronic structure evolves towards the formation of an energy miniband, 
reminiscent of that in quantum well superlattices, but driven by self-energy instead of tunneling. 

The central question about the scenario pictured in Fig.~\ref{fig1}(a) is whether the
magnitudes of the redshift ($\Delta_{red}$) and band width ($\Delta_{bw}$) 
is of practical significance.
To address this point, we solve Hamiltonian (\ref{eq:H1}) for a conduction band electron 
in a stack of $4.5$-monolayers ($1.35$ nm) thick CdSe NPLs with lateral side $L=10$ nm 
(see Supporting Information, SI, for details). 
Dielectric constants $\epsilon_{in}=10$ and $\epsilon_{out}=2$ are taken, 
which are typical values for the CdSe NPL/ligand system.\cite{BenchamekhPRB,AchtsteinNL}
Usual analytical expressions to calculate $\Sigma$, 
based on the method of image charges\cite{KumagaiPRB}, become impractical in stacks. 
We then calculate numerically the surface induced charge using Ref.~\cite{MovillaCPC} algorithms and codes. 
The ensuing $\Sigma$ compares very well with image charge results for a single quantum well (Fig. S2 in SI),
and provides a valid extension for coupled quantum wells. 
Representations of $\Sigma$ along the $z$ axis of stacks of NPLs are shown in Figs.~S3 and S4.
The self-energy potential is found to be quasi-additive down to small inter-NPL distances values, i.e.
it is roughly the superimposition of $\Sigma$ for two independent NPLs separated by $d$.
This validates the reasoning used to deduce Fig.~\ref{fig1}(a), and in particular the expected stabilization
of the central NPLs as compared to terminal ones.

Figs.~\ref{fig1}(b) and \ref{fig1}(c) (solid dots) show the magnitude of  $\Delta_{red}$ and $\Delta_{bw}$
as NPLs are brought together. 
Both magnitudes increase rapidly, following $1/d$ and $1/d^{3/2}$ scaling laws, respectively.
For a oleic acid ($d\approx 4$ nm\cite{JanaLANG}) they are in the 4-10 meV range, 
but reach few tens of meV for shorter ligands ($d \approx 1$ nm). 
We stress that this energy scale is in fact comparable to that of minibands in epitaxial 
quantum well superlattices,\cite{QW_book} and large enough to be experimentally observable e.g.
through tunneling spectroscopy.\cite{MilloPRL}
On a more conservative estimate, we set the ligand dielectric constant to $\epsilon_{out}=4$.
Because $\Sigma \propto (\epsilon_{in}-\epsilon_{out})/(\epsilon_{in}+\epsilon_{out})$,\cite{RodinaJETP,Delerue_book}
the energy splittings are reduced, but they are still sizable for short distances. 
One example is shown in Fig.~\ref{fig1}(c), empty dots.

\begin{figure}[h]
\includegraphics[width=8.5cm]{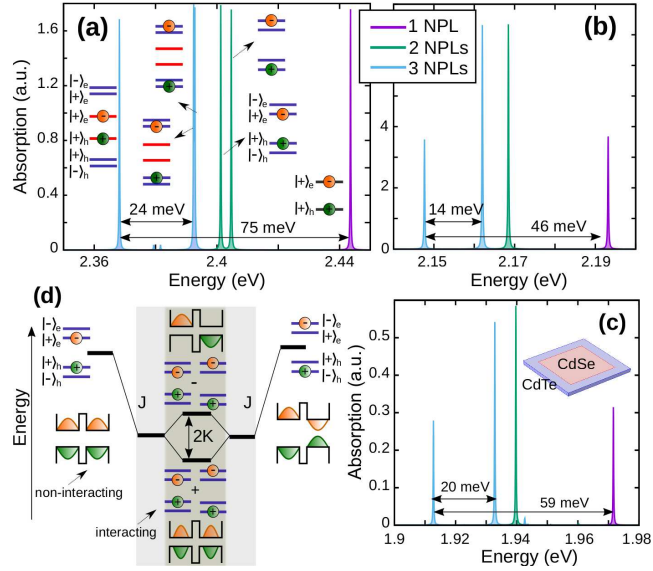}
\caption{
Absorption spectrum in stacks of 1, 2 and 3 NPLs with inter-NPL distance $d=1$ nm.
(a) e-h pair, disregarding Coulomb interaction.	
The insets show the electronic configuration of the bright states.
(b) Same but including e-h interaction (exciton).
(c) Exciton in a CdSe/CdTe core/crown NPL. 
The core (crown) side size is $L=10$ nm ($L=14$ nm).
(d) Schematic of the effect of Coulomb interaction. 
Nearly-degenerate e-h states are stabilized by $J$ and split by Coulomb tunneling to
form orthogonal (direct and indirect) exciton states.
}
\label{fig2}
\end{figure}

To investigate the optical signatures resulting from the electronic structure of Fig.~\ref{fig1}(a),
we first plot the absorption of an electron-hole (e-h) pair, disregarding Coulomb interaction. 
For convenience of presentation, we choose a small inter-NPL separation ($d=1$ nm), 
which enables a small amount of quantum tunneling.
Fig.~\ref{fig2}(a) shows the spectrum calculated for stacks of 1, 2 and 3 NPLs. 
A number of relevant features are observed: (i) the peaks redshift as the number of NPLs increases,
(ii) the number of peaks equals the number of stacked NPLs\cite{hidden}, 
(iii) the peaks of a given stack are split energetically.
Feature (i) is the consequence of the self-energy weakening in stacked NPLs, through
$\Delta_{red}$. The redshift is about twice larger than in Fig.\ref{fig1}(b) 
for the same $d$, because the self-energies of electron and hole add up.
Feature (ii) follows from selection rules. Electron and hole states can be classified by
their parity (symmetry with respect to a horizontal reflection plane in the center of the stack),
 $|\sigma_j\rangle = | + \rangle$ or $| - \rangle$.
For the e-h pair, only states with total parity even 
($|\sigma_{eh}\rangle=|\sigma_e\rangle \times |\sigma_h\rangle = | + \rangle$)
are bright.
Otherwise, electron and hole have different symmetry and their overlap is zero.
Considering one level per NPL, and the fact that parity sign alternates with
increasing energy, the only optically bright configurations are those shown in the
insets of Fig.~\ref{fig2}(a).
As for feature (iii), the origin of the energy splittings can be understood from the configurations.
For 2 NPLs, the two peaks correspond to bonding and antibonding molecular orbitals split
by tunneling energy. The splitting is small (3 meV) in spite of the short inter-NPL distance ($d=1$ nm) 
because the ligands constitute a high potential barrier.
For 3 NPLs, the second and third peak (nearly degenerate) again correspond to bonding and antibonding states 
localized in the terminal NPLs. Tunnel splitting is also present, but it is negligible because
the distance between terminal NPLs is $2d+L_z$, where $L_z$ is the thickness of the central NPL. 
The red-most peak in turn originates in the central platelet. It shows a remarkably large stabilization
($24$ meV) as compared to the peaks of terminal NPLs, this being a signature of the 
reduced $\Sigma$. 

To analyze the effect of adding e-h Coulomb interaction, we next compute exciton states from
the Hamiltonian $H_X = H_e + H_h + V_{eh}$. Here 
$V_{eh}=V_{eh}^0 + V_{eh}^{pol}$ is the e-h Coulomb attraction, with $V_{eh}^0$ describing the locally 
screened interaction and $V_{eh}^{pol}$ the interaction of one charge with the surface polarization 
created by the other one (Coulomb polarization term).\cite{RodinaJETP,Delerue_book}
Eigenstates are obtained with a full configuration interaction calculation on the basis of Hartree 
products of single-particle e-h spin-orbitals, built out of $s,\, p_x$ and $p_y$ orbitals for 
each NPL of the stack. This basis set accounts for the core contribution of both vertical and in-plane 
electronic correlations. The latter have been shown to be important in type-I NPLs.\cite{RajadellPRB,RichterPRM} 
Fig.\ref{fig2}(b) shows the resulting absorption spectrum. As compared to Fig.~\ref{fig2}(a), a few
relevant differences appear: 
(i) the whole spectrum is shifted to the red by $\sim 200$ meV, 
(ii) the interpeak splittings have shrinked (e.g. from 24 to 14 meV, in the case of 3 NPLs) and 
(iii) some peaks vanish, e.g. the second peak of the 2 NPL stack.  
The first difference is a consequence of the strong exciton binding energy.\cite{GrimNN,BenchamekhPRB,AchtsteinNL,RajadellPRB}
The second difference is because the stronger dielectric confinement of terminal NPLs 
(as compared to central ones) not only enhances $\Sigma$, but also $V_{eh}^{pol}$.
Because the two terms have opposite signs (repulsive vs attractive), there is a partial cancellation. 
It is worth noting that the cancellation is exact to first order of perturbation in strongly confined, 
spherical nanocrystals.\cite{BrusJCP} For this reason, energetic signatures of dielectric confinement 
are often negligible in superlattices of quantum dot nanocrystals.\cite{WhithamNM,JaziNL}
However, in anistropic NPLs with weak lateral confinement the compensation is far from exact, with
self-energy terms prevailing.\cite{BenchamekhPRB,PolovitsynCM,YangJPCc}
This makes colloidal NPL systems particularly suitable to tune the exciton energy through dielectric 
confinement. In addition, the compensation 
can be further reduced by resorting to type-II structures. 
Fig.\ref{fig2}(c) shows the exciton absorption in stacks of core-crown CdSe/CdTe NPL. 
Here the electron stays in the CdSe core, while the hole localizes in the CdTe crown. 
$V_{eh}$ is then reduced and, consequently, the energy splittings between peaks increase as compared
to CdSe core-only NPLs of Fig.~\ref{fig2}(b). 

The missing peaks in Fig.~\ref{fig2}(b), difference (iii), are a consequence of Coulomb tunneling.
A non-interacting e-h pair inside a pair of dielectrically equivalent NPLs gives two optically bright states, 
$|+\rangle_e \, | + \rangle_h$ and $|-\rangle_e \, | - \rangle_h$, 
possibly split by a small quantum tunneling energy (see e.g. the case of 2 NPLs in Fig.~\ref{fig2}(a)). 
Upon inclusion of Coulomb interaction, both states benefit from a direct term, 
$J=\langle + |_e \langle + |_h V_{eh} | + \rangle_e | + \rangle_h \approx
\langle - |_e \langle - |_h V_{eh} | - \rangle_e | - \rangle_h$, but 
because they share the same total parity 
they are further admixed by a crossed term, 
$K=\langle + |_e \langle + |_h V_{eh} | - \rangle_e | - \rangle_h$.
 The resulting exciton eigenfunctions are the symmetric and antisymmetric linear combinations, 
$|\sigma_{eh}\rangle_\pm = |+ \rangle_\pm = 1/\sqrt{2}\, \left( |+\rangle_e |+\rangle_h \pm |-\rangle_e |-\rangle_h \right)$,
which have Coulomb expectation values $\langle V_{eh} \rangle = J \pm K$. 
The whole interaction scheme is depicted in Fig.~\ref{fig2}(d).
It is easy to show that $|\sigma_{eh}\rangle_+$ corresponds to a direct exciton, where the
hole stays in the same NPL as the electron, and $|\sigma_{eh}\rangle_-$ to the indirect one,
where the two carriers avoid each other (see insets in Fig.~\ref{fig2}(d)).
The latter is dark because e-h overlap vanishes, 
which explains the missing peaks of Fig.~\ref{fig2}(b).
The crossed interaction $K$ is a Coulomb tunneling term, akin to quantum tunneling 
in that it splits two otherwise quasi-degenerate states in two nearby nanostructures.
It is then a molecular interaction, but acting on excitons instead of single carriers.
It reflects the stabilization (destabilization) when the e-h pair is forced to localize
in the same (opposite) NPL. Coulomb and quantum tunneling are however competitive processes,
because the former suppresses wave function delocalization in between NPLs. 
Contrary to epitaxial heterostructures, where both terms can be comparable,\cite{KrennerPRL}
in colloidal nanocrystals Coulomb tunneling is largely dominant (see Figs.~S6 and S7 in SI).


\begin{figure}[h]
\includegraphics[width=8.5cm]{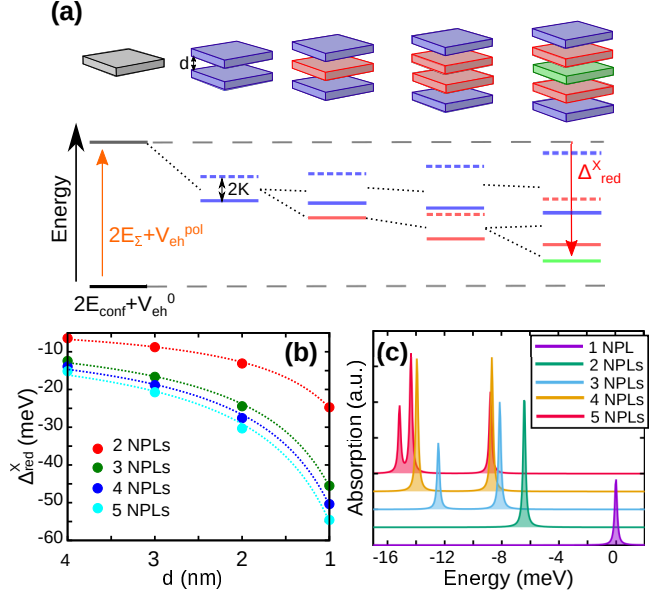}
\caption{
(a) Schematic energy spectrum of excitons with $|\sigma_{eh}\rangle=|+\rangle$ in stacks of NPLs. 
Solid (dashed) lines are direct (indirect) excitons.
(b) Exciton ground state redshift as a function of the inter-NPL distance.
Dots are calculated values, lines $1/d$ fits.
(c) Exciton absorption spectrum for $d=4$ nm, centered at the energy of one isolated NPL.
The number of peaks relates to the number of stacked NPLs as explained in panel (a).
}
\label{fig3}
\end{figure}
Gathering single particle ($\Sigma$) and two-body ($V_{eh}^{pol}$) effects
of dielectric confinement, 
we can portrait the energy structure of excitons in stacks of NPLs. 
Fig.~\ref{fig3}(a) represents a qualitative schematic, displaying exciton states with 
$|\sigma_{eh}\rangle = |+\rangle$. 
Similar to the single particle case, with increasing number of NPLs the states tend to
redshift and split owing to the weakened dielectric terms, ($2\Sigma + V_{eh}^{pol}$).
In addition, the doublets coming from a pair of equivalent NPLs split to form
direct and indirect excitons, $|+\rangle_\pm$, split by $2K$.
Only direct exciton states remain bright.
Consequently, the exciton absorption shows one bright exciton per each type of 
dielectrically equivalent NPLs, solid lines in Fig.~\ref{fig3}(a). 
This explains why the absorption spectrum, Fig.~\ref{fig3}(c), shows
a single peak for 1 or 2 NPLs, two for 3 or 4 NPLs,
and three for 5 NPLs (one for central platelet, one for terminal ones, 
one for intermediate ones). 
The same rule holds for larger number of NPLs, but the energy splittings
are increasingly small.

Fig.~\ref{fig3}(b) shows the exciton ground state redshift ($\Delta_{red}^X$) 
as a function of the inter-NPL distance, for CdSe NPLs. 
As in the single-particle case, numerical estimates are well fit by a $1/d$ scaling. 
The overall magnitude of $\Delta_{red}^X$ is similar to that of Fig.~\ref{fig1}(b),
in spite of having two particles adding their self-energies. 
This indicates that $V_{eh}^{pol}$ roughly compensates for one $\Sigma$ term. 
%

Altogether Fig.~\ref{fig3} points that it should be possible to find optical signatures of 
molecular coupling in stacked NPLs, in the form of energetically resolved multiplets of optically active peaks, 
the number of such peaks reflecting the types of dielectrically equivalent NPLs in the stack.
It is worth stressing that no such signatures are expected in the absence of dielectric 
confinement. The exciton absorption spectrum shows then a single peak, whose energy is 
independent of the number of stacked NPLs, see Fig.~S5.

One can try establish connections between the results in Fig.~\ref{fig3} and the specific 
experiments of Ref.~\cite{DirollNL}, which showed that low-temperature
emission of ensembles of presumably stacked CdSe NPLs depends on $d$.
From our theory, the exciton ground state peak of an isolated NPLs is clearly blueshifted when 
compared to that of stacked NPLs (see Figs.~\ref{fig3}(b) and (c)). 
This offers a possible explanation for the two-color emission observed in the experiments,
whereby the high-energy peak comes from isolated NPLs and the low-energy (broader) one from stacks 
with variable number of NPLs.
Different experimental features are then explained, such as the energy splitting scaling 
inversely with $d$, or the fact that the low-energy peak red-shifts and gains intensity
as the degree of stacking increases (e.g. through addition of ethanol).
However, the magnitude of the experimental splitting, $20-30$ meV, requires $d=2$ nm in our simulations, 
about half the reported inter-NPL length with oleic acid.\cite{DirollNL,JanaLANG}
This seems unlikely.
The double peak feature might still arise from exciton vs. trion emission.\cite{YuACSami,ShornikovaNL}
The $d$-dependence could then be explained from the different weight of $\Sigma$ and $V_{eh}^{pol}$ 
in exciton (neutral charge) and trion (net charge) species.
Beyond NPLs, 
the dielectric effects we predict when colloidal nanostructures are approached
provide a potential interpretation for the emission redshift recently observed in 
superlattices of CsPbBr$_3$ and charged CdTe nanocrystals, where quantum tunneling 
is not expected.\cite{RainoNAT,KimNL}
 
Further experiments are now needed to confirm the molecular coupling predicted by theory.  
Vertically stacked NPLs constitute an optimal system to this end because the reduced 
yet precisely controlled thickness permits small spacing between neighbour, isoenergetic structures.
The strongly anistropic geometry of individual NPLs also reduces compensations between 
$\Sigma$ and $V_{eh}^{pol}$, thus favoring optical manifestations.
Other ideal experimental conditions involve short ligands to minimize $d$, 
yet posing a large dielectric mismatch with the inorganic NPLs.
Large NPL lateral dimensions and reduced interlayer misorientation are convenient to 
favor degeneracy between stacked NPLs. The use of type-II hetero-NPLs to reduce 
the compensation of $\Sigma$ and $V_{eh}^{pol}$ is also benefitial, 
even though the irregular core-shell interface may constitute a source of linewidth broadening.
Low temperature absorption is desirable to avoid phonon-broadening\cite{ScottNS} 
and Forster resonant energy transfer towards defective platelets\cite{GuzelturkACS}, 
which could impair the observation of fine structure molecular effects. 
Likewise, single particle (optical or transport) spectroscopy is likely needed to 
avoid convoluted signals coming from stacks of different length.
Verifying electronic coupling between NPLs would not only prove a novel form of molecular 
interaction between nanocrystals, but also path the way to the development
of NPL superstructures whose collective properties differ from those of individual
components.

In conclusion, we have shown that the electronic structure of electrons, holes and
excitons in colloidal NPLs can be modified through stacking. 
Carriers confined in a NPL polarize the nearby NPLs, altering the dielectric environment.
This constitutes a form of molecular coupling, where the energy levels of individual
components split under the effect of self-energy, Coulomb polarization and Coulomb tunneling terms.
For large stacks, this leads to the formation of minibands with potential band width of several meV,
whose practical realization would add a degree of freedom in the design of the electronic structure,
and possibly combine the excellent optical properties of NPLs with the improved transversal
mobility of quantum well superlattices.


\section{Associated Content}

{\bf Supporting Information.} 
Includes 
(i) additional details on theoretical model, including material parameters and self-energy potential 
profiles in stacked NPLs, (ii) supporting calculations on the role of dielectric mismatch and
Coulomb tunneling in defining the optical spectrum of excitons, and wave function localization.

\acknowledgement
We acknowledge support from MICINN project CTQ2017-83781-P and UJI project B2017-59.


\begin{thebibliography}{40}
\bibitem{hidden}
The spectral feature at $2.38$ eV in Fig.~\ref{fig2}(a) is a nearly degenerate doublet.
\end{thebibliography}
\end{document}